\documentclass{article}
\usepackage{spconf,amsmath,graphicx}
\usepackage{amsfonts}
\usepackage{mathtools}
\usepackage{url}

\usepackage{makeidx}
\usepackage{amssymb}
\usepackage{dsfont}
\usepackage{amsthm}
\usepackage{hyperref}

\usepackage{tikz}
\usetikzlibrary{positioning}
\usetikzlibrary{decorations.pathreplacing}

\makeatletter
\newenvironment{leftalign*}[1][\parindent]{\setlength\hangindent{#1}\start@align\tw@\st@rredtrue\m@ne}{\endalign}

\makeatother

\def\x{{\mathbf x}}
\def\y{{\mathbf y}}

\newcommand{\R}{\mathbb{R}}
\newcommand{\manifold}{\mathcal{M}}
\newcommand{\liegroup}{\mathcal{G}}

\title{Differential Invariants for SE(2)-Equivariant Networks}
\name{Mateus Sangalli, Samy Blusseau, Santiago Velasco-Forero, Jes\'us Angulo}
\address{Centre for Mathematical Morphology, Mines Paris, PSL University}
\begin{document}
%
\maketitle
\begin{abstract}
    Symmetry is present in many tasks in computer vision, where the same class of objects can appear transformed, e.g. rotated due to different camera orientations, or scaled due to perspective. The knowledge of such symmetries in data coupled with equivariance of neural networks can improve their generalization to new samples.
    Differential invariants are equivariant operators computed from the partial derivatives of a function.
    In this paper we use differential invariants to define equivariant operators that form the layers of an equivariant neural network.
    Specifically, we derive invariants of the Special Euclidean Group SE(2), composed of rotations and translations, and apply them to construct a SE(2)-equivariant network, called SE(2) Differential Invariants Network (SE2DINNet). The network is subsequently tested in classification tasks which require a degree of equivariance or invariance to rotations. The results compare positively with the state-of-the-art, even though the proposed SE2DINNet has far less parameters than the compared models.
\end{abstract}
\section{Introduction}
\label{sec:intro}

Transformation equivariance can be introduced to a neural network to make use of the symmetry intrinsic to data features in many deep learning tasks.
Convolutional Neural Networks (CNN) for example, reduce the number of parameters and improve the results in computer vision compared to densely connected neural networks thanks to benefiting from the translation symmetry present in computer vision tasks.
Since their introduction as a general framework~\cite{cohen2016group}, group equivariant networks have been extended to work with other transformation groups, most notably translations combined with rotations~\cite{cohen2016group,worrall2017harmonic,shen2020pdo,marcos2017rotation,weiler2018learning,weiler2019general} or scalings~\cite{lindeberg2021scale}.

Many approaches to group equivariant neural networks involve first lifting the data to a higher dimensional space obtained by sampling elements of the group and performing group convolutions there \cite{cohen2016group,kondor2018generalization}. This effectively introduces equivariance to the network, but it has the downside of increasing the size and computational cost according to the group dimension, as well as requiring the group transformations to be discretized. 
 In this paper we propose instead to work with the data in the original space and to apply only equivariant operators in that space, specifically. Specifically, we apply differential invariants to the data.
 The method of \emph{moving frames} \cite{fels1998moving,olver2007generating} is used to derive fundamental differential invariants of the Special Euclidean group $\mathrm{SE}(2) \cong SO(2) \ltimes \R^2$, composed of translations and continuous rotations. The resulting networks are $\mathrm{SE}(2)$-equivariant.
 
 The rest of the paper is organized as follows. We highlight in Section~\ref{sec:related} how the present work relates to existing propositions in the literature. In Section \ref{sec:differential_invariants}, we review the method of moving frames for obtaining differential invariants and study the case of $\mathrm{SE}(2)$. In Section \ref{sec:SE2DINet}, we define the $\mathrm{SE}(2)$ differential invariants networks before testing their invariance in Section \ref{sec:exp}. We end in Section \ref{sec:conc} with some concluding remarks and future perspectives.

\section{Related Work}
\label{sec:related}
Many approaches in the literature are based on lifting the data to a higher dimensional space \cite{cohen2016group, marcos2017rotation, weiler2018learning, weiler2019general, shen2020pdo} using discretized group elements.
In this paper, the equivariant operations are performed on the same space as the input image. Moreover, we achieve equivariance to continuous rotations. In the literature, equivariance to continuous rotations was achieved using circular harmonics \cite{worrall2017harmonic} but in contrast to our method, it needs the frequencies of the feature maps to remain entangled.

In computer vision, differential invariants form a basis for many of the filtering schemes, specially in the theory of scale-spaces \cite{weickert1998anisotropic,heijmans2002morphological}. They have also been applied to the design of feature descriptors \cite{tuznik2018affine} and for the automatic learning of PDEs for image restoration \cite{liu2010learning}. Moreover, convolutional neural networks have been compared with numerical schemes for solving PDEs \cite{ruthotto2020deep} and this idea was applied to the construction of equivariant neural networks \cite{shen2020pdo} albeit not with the use of differential invariants.

\section{Moving Frames and Differential Invariants}
\label{sec:differential_invariants}
\subsection{The Method of Moving Frames}

\textbf{Moving frame.} The method of moving frames is a way of deriving differential invariants of an $m$-dimensional manifold $\manifold$ under the action of a $r$-dimensional Lie group $\liegroup$.
We denote the action of $g \in \liegroup$ on $z \in \manifold$ as $g \cdot z$.
Throughout this paper, we can assume $\manifold = \R^m$.
Note that a grayscale image with spatial support $\Omega \subseteq \R^2$, denoted by a function $u :\Omega  \to \R$, can be identified by its graph which is a surface in $\manifold = \R^2 \times \R = \R^3$ with coordinate system $(x,y,u(x,y))$.
Recall that an operator $\psi:\manifold_1 \to \manifold_2$ is $\liegroup$-equivariant if there exists actions $\pi_g, \pi^\prime_g$ on $\manifold_1, \manifold_2$ such that $\psi \pi_g = \pi^\prime_g \psi$.
A moving frame \cite{fels1998moving} is a $\liegroup$-equivariant  function $\rho : \manifold \to \liegroup$, specifically  $\rho(g \cdot z) = \rho(z) \cdot g^{-1}$ for all $g \in \liegroup$, $z \in \manifold$. This implies $\rho(g \cdot z) \cdot g \cdot z = \rho(z) \cdot g^{-1} \cdot g \cdot z = \rho(z)\cdot z$.

\textbf{Operator invariantization.} From a moving frame one can obtain an invariant operator from an arbitrary differentiable operator.
We use $\imath$ to denote the \emph{invariantization} of an operator $F:\manifold \to \R$ defined as $\imath [F](z) \coloneqq F(\rho(z) \cdot z)$. The operator $\imath[F]$ is $\liegroup$-invariant, \emph{i.e.} $\imath[F](g \cdot z) = \imath[F](z)$.

\textbf{Cross-section.} A necessary and sufficient condition for a moving frame to exist is for $\liegroup$ to act \emph{freely} and \emph{regularly}\footnote{We say that a group action is free if the only group element that fixes any
point $z$ in $\manifold$ is the identity. It is regular if the group orbits form a regular foliation on $\manifold$.} on $\manifold$ 
\cite{fels1998moving}.
In that case, a moving frame can be obtained by first defining a \emph{cross-section}.
A cross-section $K\subset \manifold$ is a restriction of the form $K = \{z \in \manifold \: | \: z_{k_1} = c_1, z_{k_2} = c_2, \dots, z_{k_r} = c_r \}$ with $k_1,\dots k_r \in \{1,\dots,m\}$ and $c_1, \dots, c_r \in \R$ are some constants.
Assuming freeness and regularity of the group action, given a cross-section $K$ the moving frame $\rho$ at point $z \in \manifold$ is the unique $g = \rho(z)$ such that $g \cdot z \in K$.

\textbf{Jet space.}  In general, the action on $\manifold$ may not be free and regular. Indeed a necessary condition is for the dimension of the orbits of $\liegroup$ to be $\dim \liegroup = r$. In the case where it is not satisfied, a solution might be to consider the higher dimensional jet space $J^n(\manifold)$, equipped with the prolonged group action. 
In the particular case where $\manifold = X \times U \cong \R^p \times \R^q$ and where we view the graph of a function $u:\R^p \to \R^q$ as a submanifold, then $J^n(\manifold) = X \times U^{(n)} \cong \R^p \times \R^{q {p+n \choose n}}$ \cite{olver1995equivalence}, and a function $u^{(n)}$ is viewed as a submanifold containing the derivatives of $u$ up to order $n$.
A \emph{differential invariant} is the invariantization of an operator $F:J^n(\manifold) \to \R$, $I = \imath[F]$.

\textbf{Image jet space.} For example, in the case where images are viewed as two-dimensional surfaces in $\R^3$ the second-order jet space $J^2(\manifold)$, for instance, is an Euclidean space of dimension $2 + {4 \choose 2} = 8$ and its elements are written as $z^{(2)} = (x, y, u, u_x, u_y, u_{xx}, u_{xy}, u_{yy})$.
The prolonged group action is the effect produced on the partial derivatives by the change in variables caused by the transformation.

\textbf{Fundamental invariants.} An important result \cite{olver2007generating} shows that the invariantizations of the partial derivatives of $u$ up to order $n$ can generate all other invariants of order $n$, and along with the invariantizations of the total derivative operators, it can generate all differential invariants. These invariants are referred to as \emph{fundamental invariants}. In the next subsection fundamental invariants of $\mathrm{SE}(2)$ are shown.

\subsection{Differential Invariants of $\mathrm{SE}(2)$ on Images}
Let us view a 2D image as a surface in $\manifold = \R^3$ given by the graph of a function $u:\Omega \subset \R^2 \to \R$ with open $\Omega$.
We seek equivariance to the special Euclidean group, $\mathrm{SE}(2)$, the group planar rotations and translations. We denote $g \in \mathrm{SE}(2)$ as $g = (R_\theta, \mathbf{v})$ where $R_\theta \in \mathrm{SO}(2)$ is a planar rotation and $\mathbf{v} = (v_1,v_2) \in \R^2$ is a translation.
It acts on $\manifold$ by
\begin{equation}
    \label{eq:se2_action}
    (R_\theta, \mathbf{v}) \cdot (x, y, u) = (\tilde{x}, \tilde{y}, u),
\end{equation}
where $\tilde{x} = x \cos \theta - y \sin \theta + v_1$, $\tilde{y} = x \sin \theta + y \cos \theta + v_2$.

The action is prolonged to the jet space $J^n(\manifold)$ in order to have a free and regular action. The prolonged action can be computed by the derivatives of $\tilde{u} = u$ with respect to $\tilde{x}$ and $\tilde{y}$.
For example, the action on the first order jet-space $J^1(\manifold)$ is $(R_\theta, \mathbf{v}) \cdot (x, y, u, u_x, u_y) = (\tilde{x}, \tilde{y}, \tilde{u},\tilde{u}_x, \tilde{u}_y)$, where $\tilde{x}$, $\tilde{y}$ are the same as in \eqref{eq:se2_action}, $\tilde{u} = u$ and 
\begin{align*}
     \tilde{u}_x  =  u_x \cos \theta - u_y \sin \theta, \quad
     \tilde{u}_y  =  u_x \sin \theta + u_y \cos \theta.
\end{align*} 
The action on $J^n(\manifold), n\geq 2$ can be derived similarly and extends the previous result, \emph{i.e.} the first five components are $(\tilde{x}, \tilde{y}, \tilde{u},\tilde{u}_x, \tilde{u}_y)$ as computed here above. By choosing an appropriate cross-section,  we compute a moving frame $\rho$ on $J^n(\manifold)$ for $n \geq 1$ from these five components only. Let $K$ be the cross-section in $J^n(\manifold)$ defined by $u_y = x = y = 0$ and $u_x > 0$, near a point $z^{(n)}$, such that $\nabla u = (u_x, u_y) \neq \vec{0}$.
Then, noting $\rho (z^{(n)})=(R_\theta, \mathbf{v})$, the constraint $\rho (z^{(n)})\cdot z^{(n)} \in K$ has a solution
\begin{equation*}
    0 = \tilde{u}_y = u_x \sin \theta + u_y \cos \theta \implies \tan \theta = \left( -\frac{u_y}{u_x} \right)  
\end{equation*}
i.e. $\theta$ is the opposite of the argument of $\nabla u$.
Moreover, $v_1 = -x \cos \theta + y \sin \theta$ and $v_2 = -x \sin \theta - y \cos \theta$. Therefore, the moving frame that solves $\rho(z^{(n)}) \in K$ is
\begin{equation}
    \rho (z^{(n)}) = \left( \frac{1}{\lVert \nabla u \rVert} \begin{bmatrix} u_x & u_y \\ -u_y & u_x \end{bmatrix},  \frac{-1}{\lVert \nabla u \rVert} \begin{bmatrix} u_x & u_y \\ -u_y & u_x \end{bmatrix} \begin{bmatrix} x \\ y \end{bmatrix}\right).
\end{equation}
This completely determines $\rho$ and we can obtain differential invariants by invariantizing the derivatives of $u$.

We denote $I_{ij} \coloneqq \imath[u_{x^i y^j}]$. Also, note $\imath[x] = \imath[y] = 0$. Then it is possible to show~\cite{olver2007generating} that given any differential operator $F(z^{(n)})$, its invariantization is given by 
\begin{equation*}
    \imath[F](x,y,u,\dots,u_{x^0 y^n}) = F(0,0,I_{00}, \dots, I_{0,n}).
\end{equation*}
Hence, if we know the forms of the invariantizations of the partial derivatives up to order $n$, we can obtain any other differential invariant of order $n$ by functional combination.
Some differential invariants with $n \leq 2$ are shown in Table \ref{tab:se2_diff_invs}.

\begin{table}[th]
    \centering
    \begin{tabular}{|c|c|}
        \hline
        $\imath[u]$ & $I_{00} = u$ \\ 
        \hline
        $\imath[u_x]$ & $I_{10} = \sqrt{u_x^2 + u_y^2}$ \\
        \hline
        $\imath[u_{xx}]$ & $I_{20} = \frac{1}{\lVert \nabla u \rVert^2} (u_{xx} u_x^2 + u_{xy} u_x u_y + u_{yy} u_y^2)$ \\
        \hline
        $\imath[u_{xy}]$ & $I_{11} = \frac{1}{\lVert \nabla u \rVert^2} (u_x u_y(u_{yy} - u_{xx}) - u_{xy}(u_y^2 - u_x^2))$ \\
        \hline
        $\imath[u_{yy}]$ & $I_{02} = \frac{1}{\lVert \nabla u \rVert^2} (u_{xx} u_y^2 - u_{xy} u_x u_y + u_{yy} u_x^2)$ \\
        \hline
    \end{tabular}
    \caption{Invariantizations of the partial derivatives of $u$ up to order $2$, assuming $\lVert \nabla u \rVert \neq 0$. These determine all other differential invariants of order two by functional combination, for example, the Laplacian $\Delta u = u_{xx} + u_{yy} = I_{20} + I_{02}$.}
    \label{tab:se2_diff_invs}
\end{table}


The recurrence equations in~\cite{olver2007generating} show that all differential invariants can be written as a functional combination of $I_{00}$, $I_{02}$ and its invariant derivatives~\cite{olver2007generating}, which for $\mathrm{SE}(2)$ are
\begin{align*}
    \mathcal{D}_1 = \imath[\partial_x] \coloneqq  {\lVert \nabla u \rVert}^{-1}(u_x \partial_x + u_y \partial_y), \\ 
    \mathcal{D}_2 = \imath[\partial_y] \coloneqq  {\lVert \nabla u \rVert}^{-1}(-u_y \partial_x + u_x \partial_y)
\end{align*}
or equivalently the directional derivatives in the directions of the gradient and its perpendicular.
More specifically, the recurrence relation has as initial conditions $I_{00}$ and $I_{02}$ given in Table \ref{tab:se2_diff_invs}, and $I_{01} = 0$, $I_{10} = \mathcal{D}_1 I_{00}$, $I_{20} = \mathcal{D}_1 I_{10}$, $I_{11} = \mathcal{D}_2 I_{10}$ 
%
%
and for $i,j$ such that $i+j = n \geq 3$:
\begin{equation}
    I_{ij} = \begin{cases} - \mathcal{D}_1 I_{i-1, j}  - P_{ij}(\mathcal{I}_{n-1}) \frac{1}{I_{10}}, \quad & \text{if} \; i > 0 \\
    - \mathcal{D}_2 I_{0, j-1}  - P_{0,j}(\mathcal{I}_{n-1}) \frac{1}{I_{10}}, \quad & \text{if} \; i=0.
    \end{cases}
\end{equation}
where $P_{ij}$ are polynomials and
$\mathcal{I}_{n}$ are  invariantizations of the derivatives up to order $n$, $\mathcal{I}_n = (I_{ij})_{0 \leq i+j \leq n} = (I_{00}, I_{10}, I_{01}, \dots, I_{0n})$.

{\textit{Remark:}} Although a differential invariant $I$ is invariant w.r.t. the action of $\liegroup$ on $\manifold$, if viewed as an operator applied to images $u:\R^2 \to \R$ \emph{i.e.} $I[u](x,y) \coloneqq I(x,y,u,u_x,\cdots,u_{x^0 y^n})$ it is $\liegroup$-equivariant: $\forall \; (R_\theta, \mathbf{v}) \in \mathrm{SE}(2)$, $I[(R_\theta, \mathbf{v}) \cdot u] = (R_\theta, \mathbf{v}) \cdot I[u]$, where $(R_\theta, \mathbf{v}) \cdot u = u(\tilde{x}, \tilde{y})$, and $\tilde{x}$ and $\tilde{y}$ are the transformed coordinates in \eqref{eq:se2_action}.

\section{Differential Invariants Networks}
\label{sec:SE2DINet}
\begin{figure}
    \centering
    \begin{tikzpicture}[every node/.style={inner sep=0,outer sep=0}, scale=.8]
      \node (original) at (0,0) {\includegraphics[width=.04\textwidth]{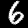}};
      \node[align=center] (label_orig) at (4.2,0) {\footnotesize input image \\ \scriptsize $H \times W \times q$};
      
      \node (deriv0) at (-2.5,-1.5) {\includegraphics[width=.04\textwidth]{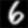}};
      \node (deriv1) at (-1.5,-1.5) {\includegraphics[width=.04\textwidth]{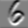}};
      \node (deriv2) at (-0.5,-1.5) {\includegraphics[width=.04\textwidth]{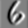}};
      \node (deriv3) at ( 0.5,-1.5) {\includegraphics[width=.04\textwidth]{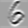}};
      \node (deriv4) at ( 1.5,-1.5) {\includegraphics[width=.04\textwidth]{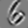}};
      \node (deriv5) at ( 2.5,-1.5) {\includegraphics[width=.04\textwidth]{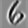}};
      
      \node[align=center] (label_deriv) at (4.2,-1.8) {\footnotesize Gaussian \\ \footnotesize derivatives \\ \scriptsize $H \times W \times q {n+2 \choose n}$};
      
      \node (inv0) at (-2,-4) {\includegraphics[width=.04\textwidth]{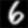}};
      \node (inv1) at (-1,-4) {\includegraphics[width=.04\textwidth]{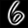}};
      \node (inv2) at ( 0,-4) {\includegraphics[width=.04\textwidth]{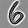}};
      \node (inv3) at ( 1,-4) {\includegraphics[width=.04\textwidth]{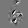}};
      \node (inv4) at ( 2,-4) {\includegraphics[width=.04\textwidth]{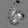}};
      
      \node[align=center] (label_deriv) at (4.2,-4.2) {\footnotesize differential \\ \footnotesize invariants \\ \scriptsize $H \times W \times q \left( {n+2 \choose n} - 1 \right)$};
      
      \node (fmap0) at (-3,-7) {\includegraphics[width=.04\textwidth]{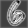}};
      \node (fmap1) at (-2,-7) {\includegraphics[width=.04\textwidth]{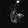}};
      \node (fmap2) at (-1,-7) {\includegraphics[width=.04\textwidth]{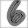}};
      \node (fmap3) at ( 0,-7) {\includegraphics[width=.04\textwidth]{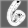}};
      \node (fmap4) at ( 1,-7) {\includegraphics[width=.04\textwidth]{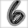}};
      \node (fmap5) at ( 2,-7) {\includegraphics[width=.04\textwidth]{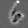}};
      \node (fmap6) at ( 3,-7) {\Large $\cdots$};
      
      \node[align=center] (label_deriv) at (4.3,-7) {\footnotesize feature maps \\ \scriptsize $H \times W \times k$};

      \draw[->] (original.south) to (deriv0.north);
      \draw[->] (original.south) to (deriv1.north);
      \draw[->] (original.south) to (deriv2.north);
      \draw[->] (original.south) to (deriv3.north);
      \draw[->] (original.south) to (deriv4.north);by the chain rule.
      \draw[->] (original.south) to (deriv5.north);
      
      \foreach \x in {0,...,5}
        \foreach \y in {0,...,4}
        {
            \draw[->] (deriv\x.south) to (inv\y.north);
        }
        
      \node (etc) at (0,-5.5) {\huge $\cdots$};
      \draw[->, ultra thick] (inv2.south) to node [right, align=center] {\scriptsize $1 \times 1$ convolutions} (etc);
      \draw[->, ultra thick] (etc) to node [right] {\scriptsize $1 \times 1$ convolutions} (fmap3.north);
      
    \end{tikzpicture}
    \caption{Illustration of a SE2DIN block applied to a MNIST-Rot image. Convolutions are followed by batch normalization and activation function. It takes as input an image of height $H$, width $W$ and $q$ channels and outputs an image with the same spatial dimensions and $k$ channels.}
    \label{fig:example_block}
\end{figure}
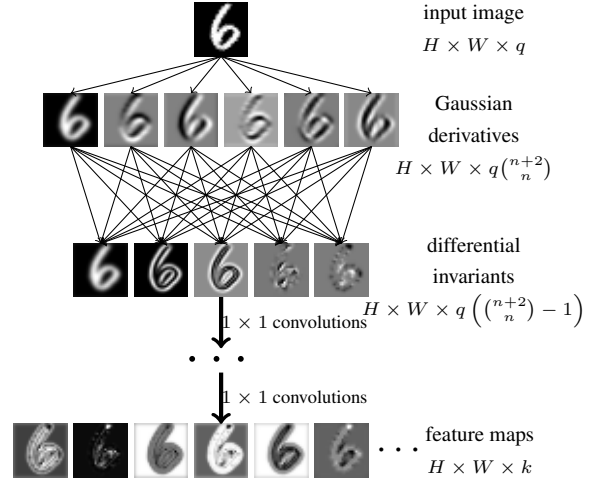

In this section we define the $\mathrm{SE}(2)$ Differential Invariants Networks (SE2DINNet). 
A SE2DINNet is made of equivariant blocks (SE2DIN blocks) consisting in computing the derivatives, followed by computing the differential invariants of $\mathrm{SE}(2)$ up to a certain order, followed by a series of $1 \times 1$ convolutions, in the style of Network in Network (NiN) \cite{lin2013network}.
We illustrate the SE2DIN block in Figure \ref{fig:example_block} and we will explain its components in this section.

\subsection{Gaussian Derivative Layers}
In order to compute the differential invariants we use Gaussian derivatives. Gaussian derivatives have already been used in neural networks to produce structured receptive fields \cite{jacobsen2016structured} in CNNs and they provide a regularization effect to the CNN.
Gaussian derivatives are used to compute the derivatives of a Gaussian filtered image, \emph{i.e.} $\frac{\partial^{i+j}}{\partial x^i \partial y^j}(u * G_\sigma) = u * \frac{\partial^{i+j}}{\partial x^i \partial y^j} G_\sigma = u * G^{i,j}_\sigma$, where $G^{i,j}_\sigma = \frac{\partial^{i+j} G_\sigma}{\partial x^i \partial y^j}$.

Gaussian filters are already $\mathrm{SE(2)}$-equivariant, so their composition with a differential invariant yields a $\mathrm{SE(2)}$-equivariant operator. These properties motivate the use of Gaussian derivatives as the first part of each SE2DIN block.

\subsection{Functional Combination}
From the derivatives, we compute the invariantizations of the derivatives up to order $n$, $\mathcal{I}_n$.
As any differential invariant $I$ of order $n$ is a functional combination $I = F(\mathcal{I}_n)$ \cite{olver2007generating}, a neural network with one hidden layer and taking as input $\mathcal{I}_n$, can approximate any differential invariant of order $n$, according to the Theorem of Universal Approximation. In practice we compute a $l$-layer densely connected network, with $l \geq 2$, applied at every spatial point $(x,y) \in \Omega$ which takes as input the vector $\mathcal{I}_n$.
Computationally this is equivalent to a series of $l$ $1 \times 1$ convolutions with pointwise activation functions applied to the multi-channel image with values $A(x,y) = \mathcal{I}_n(x,y)$.

As the moving frame was derived assuming $\nabla u \neq \vec{0}$, $I_{ij}$ is not defined for some $i,j$ when $\nabla u = \vec{0}$, therefore we introduce $\bar{I}_{ij} = \lVert \nabla u \rVert I_{ij}$ for $i + j \geq 1$, which is used instead. The limits of $\bar{I}_{ij}(x,y)$ for $(x,y) \rightarrow (x_0, y_0)$ such that $\nabla u(x_0, y_0) = \vec{0}$ are well defined for the new invariants.

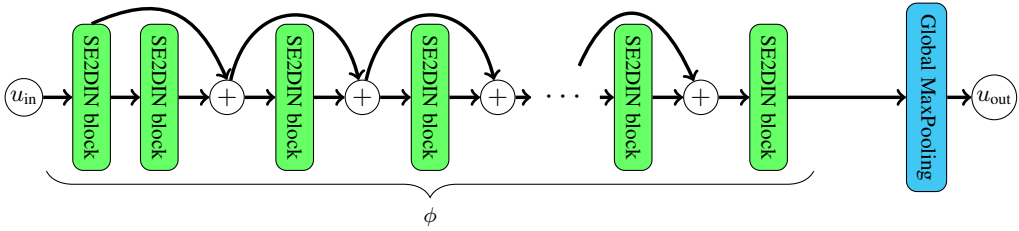
\begin{figure*}[!h]
    \centering
    \begin{tikzpicture}[scale=0.6]

		\node[circle, draw, inner sep=1pt] (input) at (0, 0) {\small $u_{\text{in}}$};		
		\node[rectangle, rounded corners, rotate=270, draw, minimum width = 1.5cm, minimum height = 0.5cm, fill=green!60,inner sep=4pt] (layer0) at (1.5, 0) {\footnotesize SE2DIN block};	
		
		\draw[line width=1.3pt,->] (input) -- (layer0);
		
		\node[rectangle, rounded corners, rotate=270, draw, minimum width = 1.5cm, minimum height = 0.5cm, fill=green!60,inner sep=4pt] (layer1) at (3, 0) {\footnotesize SE2DIN block};	
		\node[circle, draw, inner sep=1pt] (oplus1) at (4.5, 0) {$+$};		
		
		\draw[line width=1.3pt,->] (layer0) -- (layer1);
		\draw[line width=1.3pt,->] (layer1) -- (oplus1);
		\draw[line width=1.3pt,->] (layer0.west)..controls(3,2.3) and (3.8,2)..(oplus1.north);
		
	    
		\node[rectangle, rotate=270, rounded corners, draw, minimum width = 1.5cm, minimum height = 0.5cm, fill=green!60,inner sep=4pt] (layer2) at (6, 0) {\footnotesize SE2DIN block};	
		\node[circle, draw, inner sep=1pt] (oplus2) at (7.5, 0) {$+$};		
		
		\draw[line width=1.3pt,->] (oplus1) -- (layer2);
		\draw[line width=1.3pt,->] (layer2) -- (oplus2);
		\draw[line width=1.3pt,->] (oplus1)..controls(5,2.3) and (7,2.3)..(oplus2);
		
		
		\node[rectangle, rotate=270, rounded corners, draw, minimum width = 1.5cm, minimum height = 0.5cm, fill=green!60,inner sep=4pt] (layer3) at (9, 0) {\footnotesize SE2DIN block};	
		\node[circle, draw, inner sep=1pt] (oplus3) at (10.5, 0) {$+$};		
		
		\draw[line width=1.3pt,->] (oplus2) -- (layer3);
		\draw[line width=1.3pt,->] (layer3) -- (oplus3);
		\draw[line width=1.3pt,->] (oplus2)..controls(8,2.3) and (10,2.3)..(oplus3);
		
		
		\node[circle] (dots) at (12, 0) {\large $\cdots$};
		\draw[line width=1.3pt,->] (oplus3) -- (dots);
		
		\node[rectangle, rounded corners, rotate=270, draw, minimum width = 1.5cm, minimum height = 0.5cm, fill=green!60,inner sep=4pt] (layerl) at (13.5, 0) {\footnotesize SE2DIN block};	
		\node[circle, draw, inner sep=1pt] (oplusl) at (15, 0) {$+$};		
		
		\draw[line width=1.3pt,->] (dots) -- (layerl);
		\draw[line width=1.3pt,->] (layerl) -- (oplusl);
		\draw[line width=1.3pt,->] (dots)..controls(13,2.3) and (14,2.3)..(oplusl);
		

		\node[rectangle, rounded corners, rotate=270, draw, minimum width = 1.5cm, minimum height = 0.5cm, fill=green!60,inner sep=4pt] (last) at (16.5, 0) {\footnotesize SE2DIN block};	
		
		\draw[line width=1.3pt,->] (oplusl) -- (last);
		
	    \draw [decorate,decoration={brace,amplitude=10pt,mirror,raise=4pt},yshift=0pt] (.5,-1.4) -- (17.5,-1.4) node [black,midway,below,yshift=-1cm,anchor=south] {\footnotesize $\phi$};
		
		
		
		\node[rectangle, rounded corners, rotate=270, draw, minimum width = 1.5cm, minimum height = 0.5cm, fill=cyan!60,inner sep=4pt] (pool) at (20, 0) {
		\footnotesize Global MaxPooling};	
		
		\draw[line width=1.3pt,->] (last) -- (pool);
		
		\node[circle, draw, inner sep=1pt] (out2) at (21.5, 0) {\small $u_{\text{out}}$};		
		
		\draw[line width=1.3pt,->] (pool) -- (out2);

    \end{tikzpicture}
    \caption{Illustration of a basic SE2DINNet. Global maxpooling renders the architecture invariant. All blocks except the first and last have the same number of input channels $q$ and filters $k$ and follow the structure of a numerical scheme for solving a PDE.}
    \label{fig:se2dinet_illus}
\end{figure*}

\subsection{Architecture}
Given the interpretation of SE2DIN blocks as differential invariants, a SE2DINNet is viewed as a numerical scheme to solve a time-varying PDE obtained by applying the Finite Difference method to solve $u_t = F_t(\mathcal{I}_n)$~\cite{ruthotto2020deep}.
We restrict the spatial and temporal steps to $\Delta_x = \Delta_y = \Delta_t = 1$.
An example SE2DINNet is illustrated in Figure \ref{fig:se2dinet_illus}.
The network in Figure~\ref{fig:se2dinet_illus} before global max-pooling, $\phi$ is $\mathrm{SE}(2)$-equivariant, \emph{i.e.} $\phi( (R_\theta, \mathbf{v}) \cdot u) = (R_\theta, \mathbf{v}) \cdot \phi(u)$ for $u: \R^2 \to \R^{q_0}$.

\section{Experiments}
\label{sec:exp}

\subsection{MNIST-Rot}
The MNIST-Rot dataset~\cite{larochelle2007empirical} was constructed by randomly rotating by an angle in $[\pi, -\pi)$ each image from the MNIST 12k dataset~\cite{larochelle2007empirical}, which itself is obtained by selecting $12000$ images from the MNIST dataset for training/validation and $50000$ for testing. We divide the MNIST-Rot training set into $10000$ images for training and $2000$ for validation.

We test a SE2DINNet\footnote{The code used for the experiments is available at \url{https://github.com/mateussangalli/SE2DINNets}} model with six SE2DIN blocks all with two $1 \times 1$ convolutions, both with batch normalization, the first using a ReLU activation and a dropout layer.
Finally, like in Figure~\ref{fig:se2dinet_illus}, each layer except the first and last are applied by summing over the previous layer. The first two layers use $\sigma=1$ and others $\sigma=2$. We train two architectures, with orders $n=2$ and $n=3$, and with $k=20$ and $k=15$ filters respectively, to keep a similar number of weights and operations per forward propagation. We compute the third order invariants from the invariant derivatives of the second order ones.
The change in the value of $\sigma$ is done to emulate pooling.
The last layer is a global max-pooling. Optimal drop rate and weight decay parameters were obtained by a grid search.
The experiment was done ten times. Average results are shown in Table~\ref{tab:mnist_rot_acc} along with published state-of-the-art results. Our results are close to the state-of-the-art while using far less parameters and \emph{without} data augmentation.

\begin{table}[t]
    \centering
    \scalebox{.9}{
    \begin{tabular}{|c|c|c|}
        \hline
        Method & Test error($\%$) & \#params \\
        \hline
        H-Net \cite{worrall2017harmonic} & $1.69$ & $33\mathrm{k}$\\ 
        P4CNN \cite{cohen2016group} & $2.84$ & $25\mathrm{k}$ \\
        Z2CNN \cite{cohen2016group} & $5.03$ & $22\mathrm{k}$ \\
        PDO-eConv \cite{shen2020pdo} & $1.87$ & $26\mathrm{k}$ \\
        PDO-eConv $^\dagger$ \cite{shen2020pdo} & $0.709$ & $650\mathrm{k}$\\
        RotEqNet \cite{marcos2017rotation} & $1.09$ & $100\mathrm{k}$ \\
        RotEqNet $^\ddag$ \cite{marcos2017rotation} & $1.01$ & $100\mathrm{k}$ \\
        E2CNN $^\dagger$ \cite{weiler2019general} & $0.716$ & - \\
        \textbf{E2CNN} $^\dagger$ \cite{weiler2019general} & $\mathbf{0.682}$ & - \\
        SFCNN \cite{weiler2018learning} & $0.880$ & $6.5\mathrm{M}$ \\
        SFCNN $^\dagger$ \cite{weiler2018learning} & $0.714$ & $6.5\mathrm{M}$ \\
        \hline
        SE2DINNet, order 2 (Ours) & $1.64$ & $13\mathrm{k}$\\
        SE2DINNet, order 3 (Ours) & $1.62$ & $11\mathrm{k}$ \\
        \hline
        \multicolumn{3}{|c|}{Trained on non-rotated data}\\
        \hline
        RotEqNet \cite{marcos2017rotation} & $20$ & - \\
        Z2CNN & $67.9$ & $22\mathrm{k}$ \\
        Z2CNN$^\dagger$ & $6.12$ & $22\mathrm{k}$ \\
        SE2DINNet, order 2 (Ours) & $2.55$ & $13\mathrm{k}$\\
        SE2DINNet, order 3 (Ours) & $2.49$ & $11\mathrm{k}$ \\
        \hline
    \end{tabular}
    }
    \caption{Mean accuracies on the MNIST-Rot dataset. Legend: $\dagger$ - train time augmentation, $\ddag$ - test time augmentation.}
    \label{tab:mnist_rot_acc}
\end{table}

\subsection{Training on Non-Rotated Data}
Here we keep the training set of the MNIST12K dataset, without random rotations, but use the test set of MNIST-Rot. The same SE2DINNet architectures are evaluated. We train a CNN baseline, with the Z2CNN \cite{cohen2016group} architecture, with and without data augmentation.
The results are also in Table \ref{tab:mnist_rot_acc}. As expected, these models performed worse than those trained on the full MNIST-Rot. However, compared to similar experiments \cite{marcos2017rotation} trained on $10000$ samples of the original MNIST and tested on MNIST-Rot, our proposed architectures obtain a large improvement.

%

\section{Conclusion}
\label{sec:conc}

We derived differential invariants for $\mathrm{SE}(2)$ using the moving frames method and constructed an invariant neural network architecture which use these invariants as its equivariant layers. Moreover, these networks were tested in tasks to evaluate their invariance and compared well to the state-of-art results while using far less parameters and not using data augmentation. We also observe that the third order models performed slightly better than the second order ones, which motivates investigating orders higher than three. The proposed approach can be adapted to other settings e.g. networks with other symmetry groups, for example the centro-affine~\cite{olver2010moving} and projective~\cite{olver2020projective} groups. The approach can also be applied other data that can be represented as manifolds, such as other surfaces embedded in $\R^3$.

\textbf{Acknowledgement.}
This work has been supported by FMJH under the PGMO-IRSDI 2019 program.


\bibliographystyle{IEEEbib}
\bibliography{refs}

\end{document}